\documentclass{INTERSPEECH2023}
\usepackage{url}
\usepackage{graphicx}
\usepackage{amsmath}
\usepackage{lipsum}
\usepackage[mathscr]{euscript}
\usepackage{pifont}
\usepackage{wasysym}
\usepackage{amssymb}
\usepackage{multirow}
\usepackage{hyperref}
\usepackage{pifont}

\interspeechcameraready
\title{Re-ENACT: Reinforcement Learning for Emotional Speech\\
Generation using Actor-Critic Strategy}

\name{Ravi Shankar$^1$, Archana Venkataraman$^1$}
\address{
  $^1$Department of Electrical and Computer Engineering, Johns Hopkins University, Baltimore MD, USA}
\email{rshanka3@jh.edu, archana.venkataraman@jh.edu}

\begin{document}

\maketitle
 
\begin{abstract}

\end{abstract}
In this paper, we propose the first method to modify the prosodic features of a given speech signal using actor-critic reinforcement learning strategy. Our approach uses a Bayesian framework to identify contiguous segments of importance that links segments of the given utterances to perception of emotions in humans. We train a neural network to produce the variational posterior of a collection of Bernoulli random variables; our model applies a Markov prior on it to ensure continuity. A sample from this distribution is used for downstream emotion prediction. Further, we train the neural network to predict a soft assignment over emotion categories as the target variable. In the next step, we modify the prosodic features (pitch, intensity, and rhythm) of the masked segment to increase the score of target emotion. We employ an actor-critic reinforcement learning to train the prosody modifier by discretizing the space of modifications. Further, it provides a simple solution to the problem of gradient computation through WSOLA operation for rhythm manipulation. Our experiments demonstrate that this framework changes the perceived emotion of a given speech utterance to the target. Further, we show that our unified technique is on par with state-of-the-art emotion conversion models from supervised and unsupervised domains that require pairwise training.

\section{Introduction}
Human speech carries a plethora of information beyond language and grammar. It conveys details about the speaker, their mood and intent. Reproducing or injecting these para-linguistic characteristics in machine generated speech is currently an active area of research. Emotional speech synthesis has many applications such as providing personalized customer-support, developing voice-assisted therapy for elderly, and designing human-computer interface systems~\cite{emotion_personality, voice_quality}. Modifications in pitch and energy contour can inject emotional cues into the neutral speech or change the overall speaking style~\cite{psych, diffeomorphic_hnet, hnet_max_likelihood, mellotron, vcgan}. These prosodic features are also used to evaluate the quality of human machine dialog systems~\cite{prosody_eval}, and they play a significant role in speaker identification and recognition~\cite{voice_quality}. A central idea connecting most of these recent published works is learning a transformation or mapping function for the prosodic features from one emotion to another~\cite{vcgan_journal} for synthesizing speech. 
\begin{figure}
    \centering
    \includegraphics[height=2.5cm, width=0.7\linewidth]{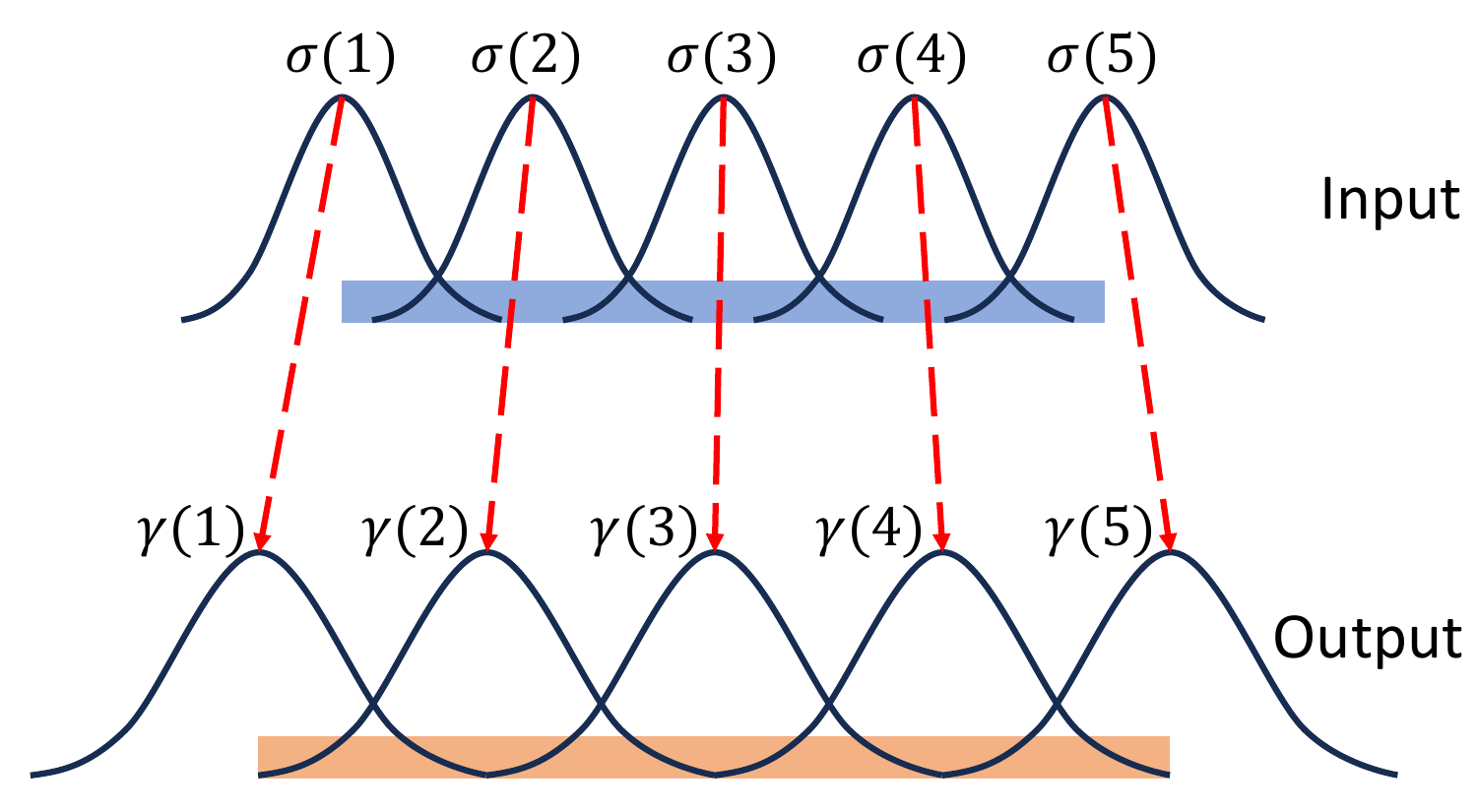}
    \caption{Overlap add operation to stretch the input signal.}
    \label{fig:wsola}
\end{figure}
Rhythm or cadence, in particular, plays a crucial role in conveying emotions~\cite{emotion_perception_factors} and in diagnosing human speech pathologies~\cite{stutter_evaluation}. Despite this, very few works on emotional speech generation/conversion target rhythm modification~\cite{adaptive_duration}. This can be attributed to the difficulty in modeling rhythm as an intrinsic parameter of speech like pitch and intensity. Further, sequence-to-sequence algorithms allows learning of the rhythm aspect without explicit modeling, but they require huge amount of good quality parallel utterances which can be expensive. 

Authors in~\cite{adaptive_duration} adopted a dynamic time warping procedure baked directly into a neural network to perform local and global rhythm modification. However, its main limitation is the supervised paradigm of learning that again requires parallel data to train the underlying generative model. In order to solve this problem in an unsupervised manner, we will make a few simplifications in the problem statement and adopt an actor-critic reinforcement learning strategy~\cite{intro_to_rl, actor_critic, policy_gradient}. We will further expand the scope of our modeling technique to the other critical prosodic features for emotion conversion, namely: pitch and intensity. To be more specific, the rhythm modification for emotion/voice conversion has three sub-problems:
\begin{itemize}
    \item \textbf{Identifying informative segments of emotion in speech}
    \item \textbf{Predicting a factor of modification for each segment}
    \item \textbf{Modifying the rhythm of these segments using WSOLA}
\end{itemize}
In the proposed approach, we will not modify the prosody of every single phoneme/syllable in an utterance. Instead, we modify only a subset of these segments (most important ones) which are identified by a Markov temporal mask for the task of emotion recognition. We adjust the prior on this temporal mask to identify segments spanning a complete syllable or a word. However, they are allowed to be a collection or mix of any of these components including short pause and silences which is a limitation in~\cite{adaptive_duration}. After identifying such segments (countable in practice), we will process them sequentially to predict a modification factor for length, pitch and intensity. These factors are uniform for the entire length of the segment. Finally, after manipulating the features by their corresponding predicted factor, we re-synthesize the utterance with a different emotional intent than before. In our knowledge, this is the first fully-unsupervised model for emotion conversion targeting pitch, intensity and rhythm in a single unified framework. Our experiments on a multispeaker corpus will demonstrate the efficacy of the proposed approach in comparison to prior techniques. 

In the rest of this paper, we will solve each of the three sub-problems sequentially. First, we will discuss the mechanism of length modification in speech (owing to its difficulty in modeling), followed by our Markov masking strategy for discovery of important segments in a weakly supervised setting. Finally, we will predict the factor of modification using actor-critic reinforcement learning strategy. Segment discovery will rely on prediction of human perception of emotional saliency using VESUS~\cite{vesus} and CREMA-D~\cite{cremad} corpora.

\begin{figure*}[t]
    \centering
    \includegraphics[height=5.3cm, width=0.9\linewidth]{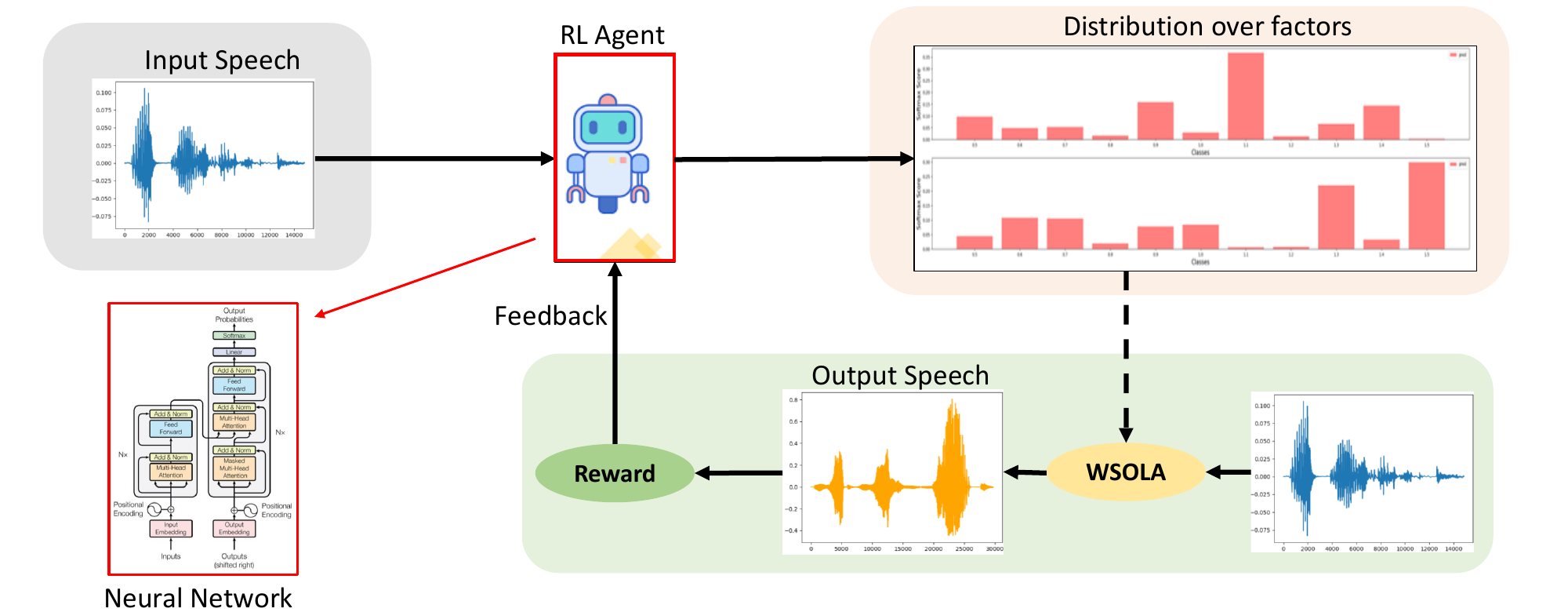}
    \caption{RL Strategy: Reinforcement learning framework for predicting factor of modification. The grey panel summarizes the state of the observer, the red panel constitutes the action space, and the green panel represents the environment with WSOLA.}
    \label{fig:rl_strategy}
\end{figure*}

\section{Method}

\subsection{Mechanism of Modification}
There are many algorithms which allow speech length modification such as overlap-add (OLA)~\cite{wsola, wsola_orig}, wave similarity overlap-add (WSOLA) and phase vocoder, to name a few. The underlying principle in these methods is to split the input speech signal into chunks of a fixed length, and use overlap add to expand or shorten the duration of signal by local replication or truncation. In~\cite{adaptive_duration}, the authors use the same local replication and deletion of frames from the input signal to modify its rhythm. While this approach works in practice, it also introduces discontinuities in the phase component. As a result of this discontinuity, the modified signal has a noticeable choppy effect for the listeners. WSOLA reduces this artifact via a correlation based search strategy to locally find the best segment for reconstruction in the neighborhood at any given frame.  

To modify duration of a signal using overlap-add, the first step is to specify the type of window function $w(n)$, its width $l$ and overlap factor $\eta$. Hanning window of width more than the lowest expected fundamental frequency is a common choice for this operation. Then, the length of output signal $z(n)$, and the overlap factor decides the time-stamps where the window's center should appear in the input signal $y(n)$. Specifically, let $\tau(n)$ be the time-stretching function, the position of window on output signal can be easily derived via:
\begin{equation}
    \gamma(1) = 1 \text{ and } \gamma(k) = \gamma(k-1) + \eta
\end{equation}
Here the total number of $\gamma$ is $\lceil \arrowvert z \arrowvert / \eta \rceil$ where $\arrowvert z \arrowvert$ denotes the length of input signal. Knowing the $\gamma(k)$, we can figure out the position of windows on input signal $y$ by $\sigma(k) = \tau^{-1}(\gamma(k))$. It is very important for the time-stretching function $\tau$ to be monotonic in nature for invertibility. Finally, the reconstruction by overlap-add algorithm is given by:
\begin{equation}
    z(n) = \frac{\sum\limits_{k=1}^{len(\sigma)} w(n - \gamma(k)) \cdot y(n - \gamma(k) + \sigma(k))}{\sum\limits_{k=1}^{len(\sigma)}w(n - \gamma(k))}
    \label{eqn:overlap_add}
\end{equation}
The choice of Hanning window with an overlap factor of $ \geq 0.5$ ensures that the denominator in Equation~\ref{eqn:overlap_add} always adds up to a constant ($1$ when equal to $0.5$). Fig.~\ref{fig:wsola} represents the schematic diagram of overlap-add operation at a high level.

The main drawback of incorporating WSOLA algorithm in any data-driven rhythm modification algorithm is its non-differentiable nature. Therefore, the loss function for optimization is infeasible due to the lack of a functional form as backpropagation is undefined. However, we can use reinforcement learning strategy where the WSOLA operation can be declared part of the agent's interaction environment to skip backpropagating through it. This is the solution we focus on in our model-free reinforcement learning for predicting modification factors. 

\subsection{Salience Prediction}
As mentioned before, estimating the salient regions for emotion perception is the second piece of the puzzle. We employ a simple masking strategy (similar to the attention maps) in order to recover a continuous segment of speech responsible for human perception of emotion~\cite{attention}. Corpora like VESUS~\cite{vesus} consists of utterances in 5 emotion categories, namely: neutral, angry, happy, sad and fearful. In addition to these audio files, each utterance in VESUS has annotations obtained from 10 listeners on Amazon Mechanical Turk (AMT) asking them to identify emotions in the corresponding utterance. The ratings provided by these listeners therefore, form a categorical distribution over the emotion classes. This is important for multiple reasons: (a) it gives us an idea of how strongly a specific emotion is portrayed, (b) the soft assignment can be used to inject confidence in the model for emotion prediction task. Therefore, our task boils down to predicting the perception score for each emotion using the content from the masked portion of input speech. 

To think simplistically, attention mechanism is a fairly straightforward approach to solve this problem. However, without any additional constraints, it may not discover any contiguous segments of speech (e.g. single frames) which are non-informative for downstream task of emotion conversion. Therefore, our objective is to find continuous segments (syllable/word level). This facilitates easy manipulation of prosodic features, i.e., pitch, intensity and rhythm by WSOLA. We devise a clever masking strategy that allows us to discover such segments. Specifically, we design a neural network with three components: (a) feature extractor module that is made entirely of downsampling convolution layers, (b) a mask generator module that estimates a collection of Bernoulli random variables over the features and (c) a salience predictor module to predict emotional saliency using information contained in the masked region. Fig.~\ref{fig:salience_network} shows the neural network architecture in detail. 

\begin{figure}
    \centering
    \includegraphics[width=\linewidth, height=3.5cm]{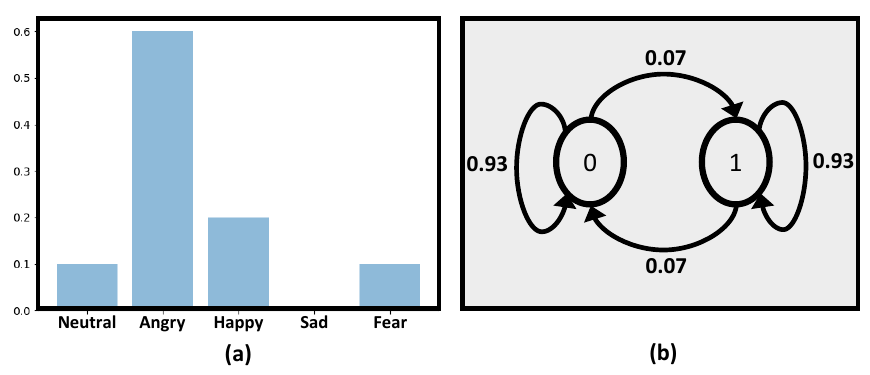}
    \caption{(a) Example of salience score obtained from AMT and (b) Transition diagram for masking random variables.}
    \label{fig:salience_and_transition}
\end{figure}

\subsubsection{Masking Variable}
\begin{figure}
    \centering
    \includegraphics[height=3.3cm, width=\linewidth]{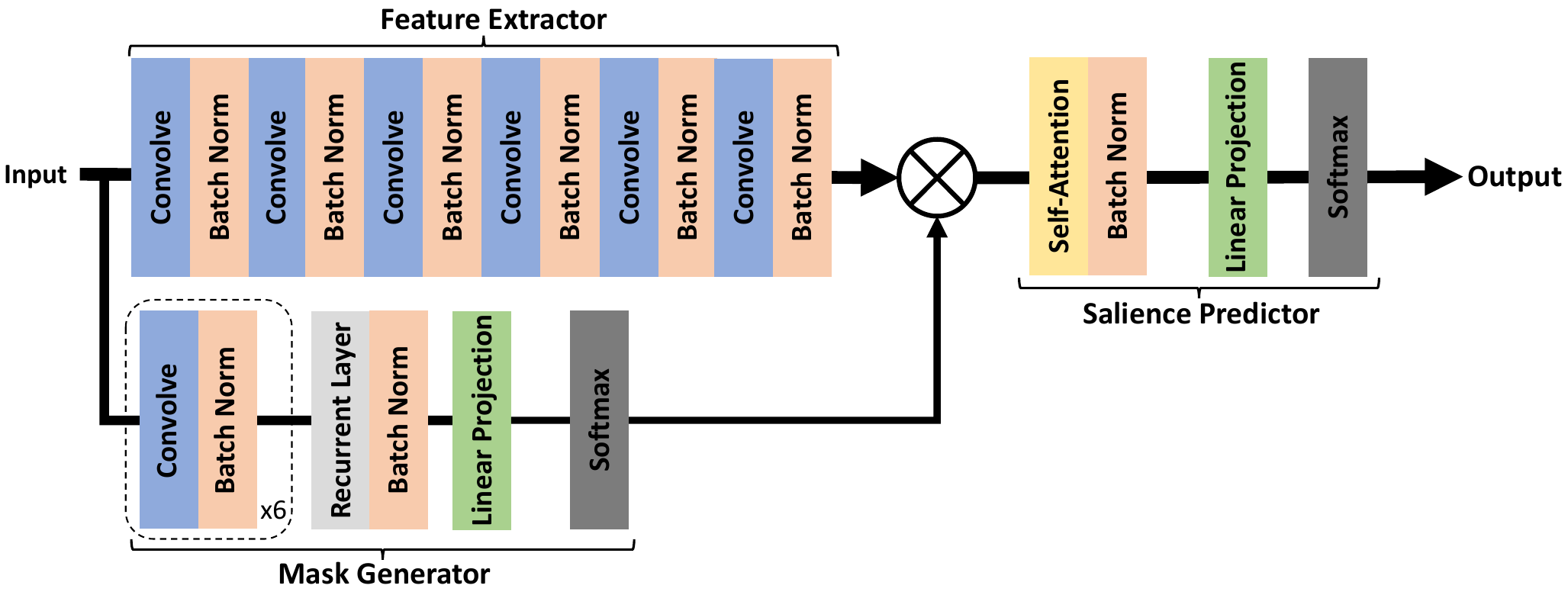}
    \caption{Neural network used for prediction of human perception of emotional saliency. The architecture has three components: (a) feature extraction from raw waveform, (b) mask generator using Markov masking and (c) salience prediction.}
    \label{fig:salience_network}
\end{figure}

We generate mask via sampling from the variational posterior learned by a combination of recurrent and linear projection layers (Fig.~\ref{fig:salience_network}). Given a sequence of frames $X_t \ \forall \ t \in [1,....T]$, the binary mask sequence $[M_1, M_2, M_3.... M_T]$ is a collection of $T$ Bernoulli random variable with the following prior: 
\begin{equation*}
P(M_t \arrowvert M_{t-1}) = 
    \begin{cases}
    Ber(p), & \text{if } M_t = M_{t-1}\\
    Ber(1-p), & \text{otherwise}
\end{cases}
\end{equation*}
$\forall \ t = 2, 3, ...T$ and $p \in (0,1)$. Further, we specify $P(M_1) = Ber(0.01)$ so that the masking follows a first-order Markov property, i.e. the future is independent of past given present and has low activation for the first frame of signal. The mask distribution as time $t$ is dependent on the mask at time $t-1$. It is by-design similar (0 or 1) to the previous time-step, ensuring a continuity in the segments. Fig.~\ref{fig:salience_and_transition}(b) shows the state transition diagram of the corresponding Markov chain. While this prior constraint on the mask helps identify continuous segments, it can happen that the mask takes the value $1$ for the entire duration of the speech utterance. Hence, we define the mean-field distribution over the mask estimated by the neural network as: 
\begin{equation}
\text{\footnotesize $q_{\theta}(M_1, ...., M_T \arrowvert \mathbf{X}) = q_\theta(M_1 \arrowvert \mathbf{X}) q_\theta(M_2 \arrowvert \mathbf{X}) \, ... \, q_\theta(M_T \arrowvert \mathbf{X})$}
\end{equation}
where, we have used the mean-field approximation~\cite{blei, murphy} for the variational posterior learned by the neural network (parameterized by $\theta$). We add a sparsity penalty at each time step via KL divergence loss with a Bernoulli distribution of very small success. Specifically, the sparsity penalty can be written as:
\begin{equation}
    \mathcal{L}_{sparse} = \sum\limits_{t=1}^{T}D_{KL}\big[ q_\theta(M_t \Arrowvert \mathbf{X}) \arrowvert Ber(0.01) \big]
\end{equation}
Adding the sparsity penalty to approximate posterior solves the problem of the mask being triggered for the entire signal's duration. Finally, the Markov prior is imposed on the posterior using KL divergence penalty which can be written as:
\begin{align*}
    & \text{\footnotesize $\mathcal{L}_{prior} = D_{KL} \big[ q_\theta(\mathbf{M} \arrowvert \mathbf{X}) \Arrowvert P(\mathbf{M}) \big]$} \\
    & \text{\footnotesize $\phantom{\mathcal{L}_{prior}} = D_{KL} \big[ q_\theta(M_1, M_2 ... M_T \arrowvert \mathbf{X}) \Arrowvert P(M_1, M_2, ... M_T) \big]$} \\
    & \text{\footnotesize $\phantom{\mathcal{L}_{prior}} = \sum_{M_T}\sum_{M_{T-1}} ... \sum_{M_1} q_\theta(M_1 \arrowvert \mathbf{X}) q_\theta(M_2 \arrowvert \mathbf{X}) ... q_\theta(M_T \arrowvert \mathbf{X})$} \\
    & \text{\footnotesize $\phantom{\mathcal{L}_{prior} \qquad \qquad} \times \log{\frac{q_\theta(M_1 \arrowvert \mathbf{X}) q_\theta(M_2 \arrowvert \mathbf{X}) ... q_\theta(M_T \arrowvert \mathbf{X})}{P(M_1)P(M_2 \arrowvert M_1) .. P(M_T \arrowvert M_{T-1})}}$} \\
    & \text{\footnotesize $\phantom{\mathcal{L}_{prior}} = \sum_{M_1} q_\theta(M_1 \arrowvert \mathbf{X}) \log{\frac{q_\theta(M_1 \arrowvert \mathbf{X})}{P(M_1)}}$} \\
    & \text{\footnotesize $\phantom{\mathcal{L}_{prior} \qquad \qquad} \times \sum_{M_2} q_\theta(M_2 \arrowvert \mathbf{X}) \log{\frac{q_\theta(M_2 \arrowvert \mathbf{X})}{P(M_2 \arrowvert M_1)}} $}\\
    & \text{\footnotesize $\phantom{\mathcal{L}_{prior} \qquad \qquad} ... \times \sum_{M_T} q_\theta(M_T \arrowvert \mathbf{X}) \log{\frac{q_\theta(M_T \arrowvert \mathbf{X})}{P(M_T \arrowvert M_{T-1})}} $}\\
\end{align*}
Therefore, the KL-divergence penalty is decomposed into $T$ terms where each term can be computed conditioned on the past which has only two values, i.e., 0 or 1. Hence, the computation is tractable and vectorized for efficiency. Further, we condition the masking prior on energy based thresholding to avoid leading and trailing silences. Note that, setting p to 0.93 results in masking probability dropping below 0.5 after 180ms which is a typical syllable duration.

\subsubsection{Neural Network for Salience Prediction}
We parameterize the distribution $q_\theta(M_t \arrowvert \mathbf{X})$ by Bernoulli parameters estimated using sigmoid activations in the mask generator to get their posteriors (Fig~\ref{fig:salience_network}). Then, we sample from approximate posterior to generate the mask which gets element-wise multiplied with the extracted features (from convolutional stack) and fed into the salience prediction module. The saliency loss is a simple L-1 penalty over the predicted softmax (5 classes: neutral, angry, happy, sad and fear) and the ground truth. Finally, since we sample from the variational posterior to generate the mask for salience prediction, we use Gumbel softmax~\cite{gumbel, gumbel_max} for backpropagation through the sampler module. Therefore, the training loss for saliency prediction is given by: 
\begin{equation}
    \mathcal{L}_{saliency} = \Arrowvert Y - \hat{Y}\Arrowvert + \lambda_{prior} \mathcal{L}_{prior} + \lambda_{sparse} \mathcal{L}_{sparse}
\end{equation}
where, $Y$ and $\hat{Y}$ are the ground-truth saliency (obtained from AMT) and saliency predicted by model in Fig.~\ref{fig:salience_network}.

\subsection{Factor of Modification: Policy Gradient}
From the last subsection, we have a strategy for obtaining portions of speech that affects our emotion perception. Knowing these allow us to use WSOLA algorithm to modify their length. Here, we will discuss our approach to get a distribution over the factors of modification using actor-critic reinforcement learning strategy. Our first step is to discretize the space of possible duration factors for simplification. We choose a range of 0.25-1.9 in steps of 0.15. Note that, this covers a very wide range for WSOLA operation. Extreme modifications in speaking rate can lead to distortions in the signal. However, we found this interval to be within the operating range. Furthermore, by creating a finite number of classes we can learn a categorical distribution. 

We employ the offline actor-critic policy gradient method to estimate the factor of modification given the speech signal and information about the salient segments~\cite{policy_grad_1, policy_grad_2, intro_rl}. Denoting the speech utterance by $y(t)$ and the mask variable by $M_t \; \in \{0,1\}$, the state of the system can be characterized by the tuple $S_\tau = (y_t, M_t)$. Mask variable $M_t$ is an indicator function (same length as $y$) denoting the important segments of the signal. The learning agent takes this state tuple and the target emotion description (as one-hot vector representation) to predict a distribution over the discrete set of factors describing the action space $A_\tau$ and the advantage of current state $S_\tau$. 

After sampling from the action space distribution, we modify the length of corresponding segment using WSOLA and obtain a reward signal $r$. This reward signal measures the goodness of predictive distribution over the actions for emotion modification, i.e., the increase in target emotion category score measured by salience predictor. To train the model efficiently, we uniformly sample one salient region at a time during training. Therefore, an episode is a single time-step. Fig.~\ref{fig:rl_strategy} shows the complete RL framework at a coarser level. We will now dive into the details of the reinforcement learning agent used here. 

\begin{figure}
    \centering
    \includegraphics[width=0.9\linewidth, height=3.cm]{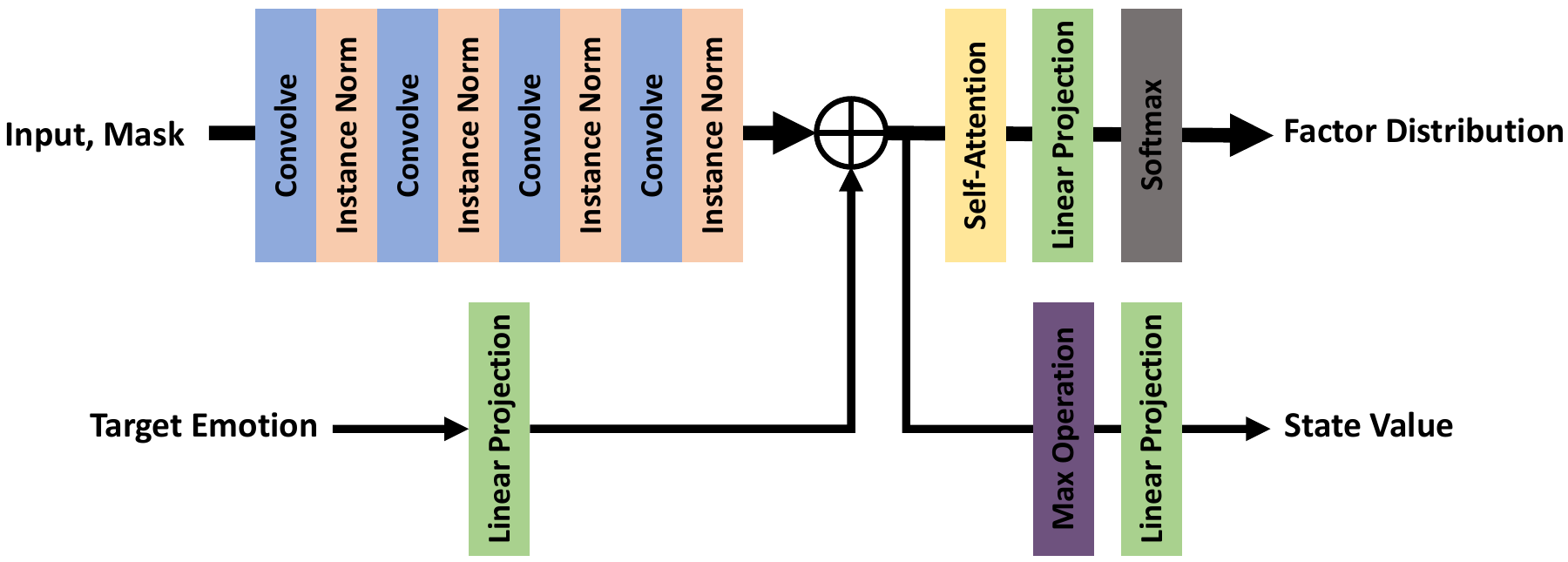}
    \caption{RL architecture used for factor prediction.}
    \label{fig:rl_agent}
\end{figure}
\subsubsection{RL Agent}
The reinforcement learning (RL) agent is a deep neural network~\cite{adamae} consisting a stack of convolution and transformer layers to learn appropriate distribution over the actions $A$. It is conditioned on three quantities: (a) the input speech signal (in time domain), (b) segment mask through indicator variables, and (c) the target emotion code corresponding to which a prediction has to be made. Fig.~\ref{fig:rl_agent} shows the neural network architecture used for estimating a probability distribution over the allowable set of factors. Since, the distribution is over an entire masked region, we use max-pooling on the output of transformer layer to feed into the final softmax layer. Therefore, the policy function is a neural network parameterized by $\theta$. The objective of this neural network is to maximize the expected reward which can be written as~\cite{policy_grad_1}:
\begin{align*}
    & \mathcal{L}(\theta) = E_{\pi} \Big[ r(s) \Big] \Rightarrow \nabla \mathcal{L}(\theta) = \nabla \sum_{a \in A} \pi(a \arrowvert s) r(s) \\
    & \nabla \mathcal{L}(\theta) = \sum_{a \in A} \nabla \pi(a \arrowvert s) r(s) \\
    & \phantom{\nabla \mathcal{L}(\theta)} = \sum_{a \in A} \pi(a \arrowvert s) \nabla \log{\pi(a \arrowvert s)} r(s) \\
    & \phantom{\nabla \mathcal{L}(\theta)} = E_{\pi} \Big[ r(s) \nabla \log{\pi(a \arrowvert s)} \Big]
\end{align*}
The objective function suggests that, we do not need estimation of gradients through the reward model for training the agent network. It is specially helpful in our case because the reward framework uses WSOLA operation which is not differentiable. 

\begin{table}[t!]
\begin{center}
\begin{tabular}[t]{ |c|c|c|c|c| } 
 \hline
 \textbf{Dataset} & \textbf{Mode} & \textbf{F1 (macro)} & \textbf{F1 (wtd)} & \textbf{Acc.}\\
 \hline
 \multirow{2}{4em}{VESUS} & Top-1 & 0.72 & 0.78 & 0.82\\ 
 & Top-2 & - & - & 0.94\\
 \hline
 \multirow{2}{4em}{CREMAD} & Top-1 & 0.59 & 0.65 & 0.66\\ 
 & Top-2 & - & - & 0.83\\
 \hline
\end{tabular}
\caption{Emotion recognition performance on multiple datasets.}
\label{tab:emo_recog}
\end{center}
\end{table}

\section{Experiments and Results}
In this section, we will discuss the results of saliency prediction using proposed technique. Our evaluation covers both objective and subjective metrics in terms of preference scores. We compare our RL-based technique to existing state-of-the-art models for supervised/unsupervised emotion conversion published recently. We start with a description of data used for experiments. 

\begin{figure}[!t]
    \centering
    \includegraphics[width=\linewidth, height=3.5cm]{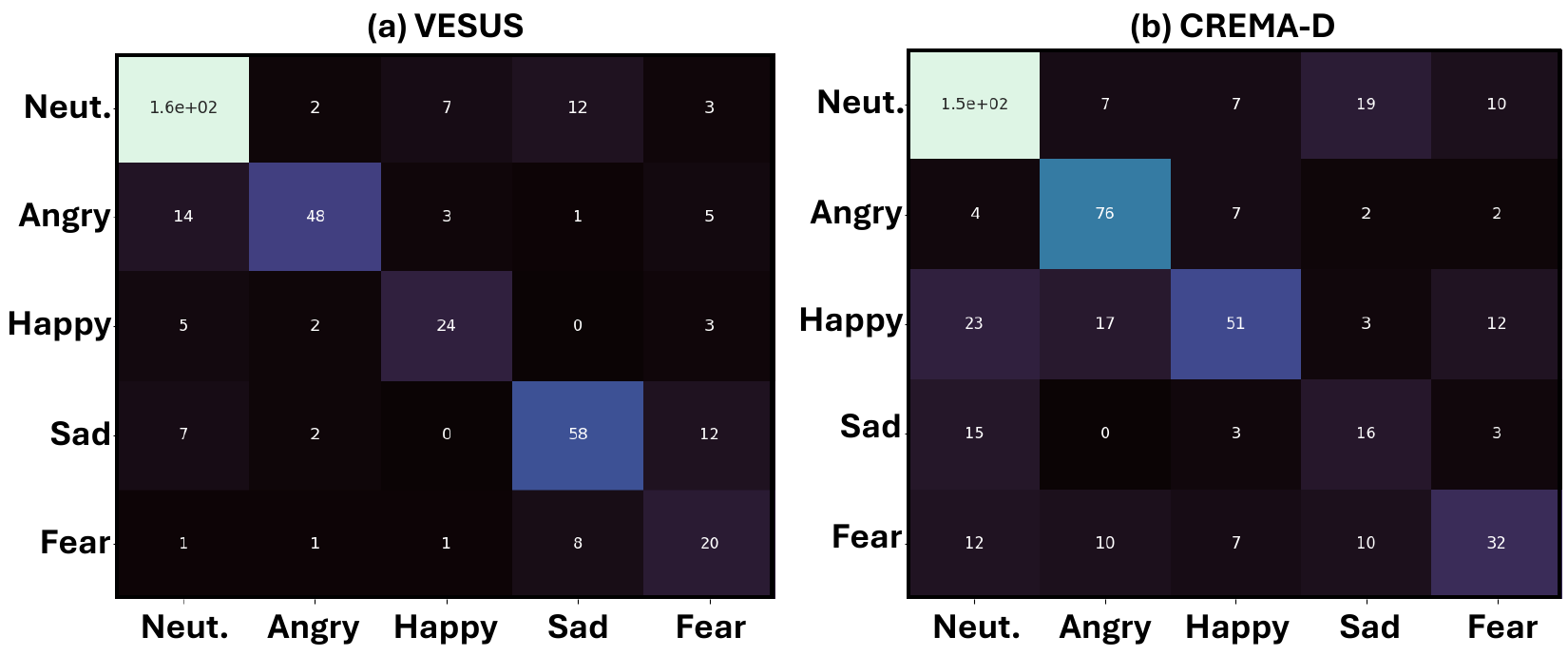}
    \caption{Confusion matrices corresponding to the top-1 accuracy on (a) VESUS and (b) CREMAD test set.}
    \label{fig:confusion_matrix}
\end{figure}

We use the VESUS~\cite{vesus} corpus to carry out our evaluations. VESUS provides a crowd-sourced annotation obtained from 10 listeners on Amazon Mechanical Turk (AMT) for each sample. This allows us to create a soft assignment over the mixture of emotion (neutral/angry/happy/sad/fearful) rather than a single emotion category for prediction. As a result, our proposed salience predictor predicts the emotion perceived by AMT listeners during training and inference stage. Further, we split the VESUS according to the following scheme:
\begin{itemize}
    \item 11.5k samples are randomly selected for training
    \item 150 samples are randomly selected for validation
    \item 400 samples are randomly chosen for evaluation/testing
\end{itemize}

\begin{figure}[!t]
    \centering
    \includegraphics[width=\linewidth, height=5.8cm]{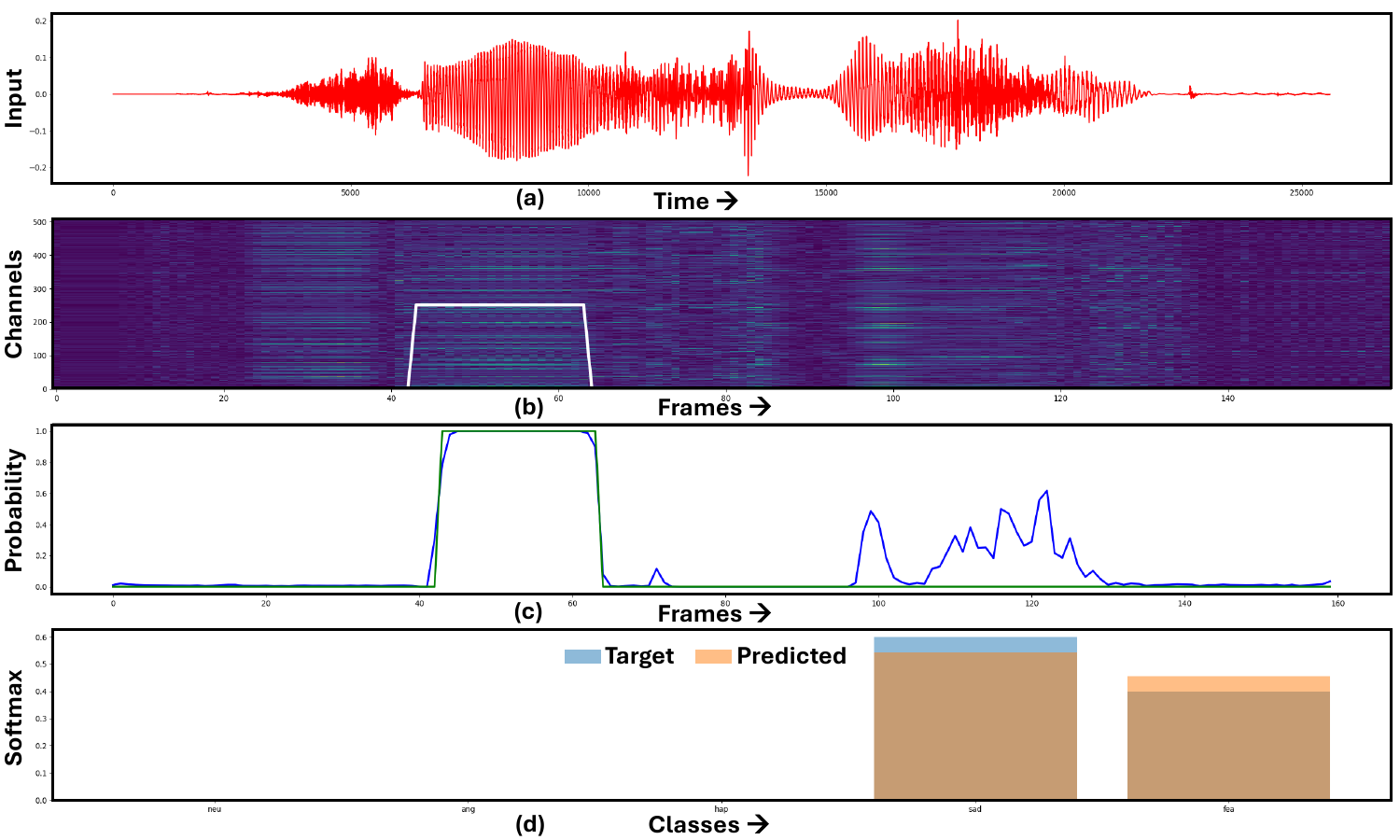}
    \caption{Example of discovered segments important for prediction of the corresponding emotion classes. (a) Time domain input signal, (b) overlaying the sampled mask on features extracted from convolution block, (c) variational posterior of the mask predictor, and (d) predicted saliency compared to target.}
    \label{fig:segments_demo}
\end{figure}

\begin{figure*}[!t]
    \centering
    \includegraphics[width=0.85\linewidth, height=4.2cm]{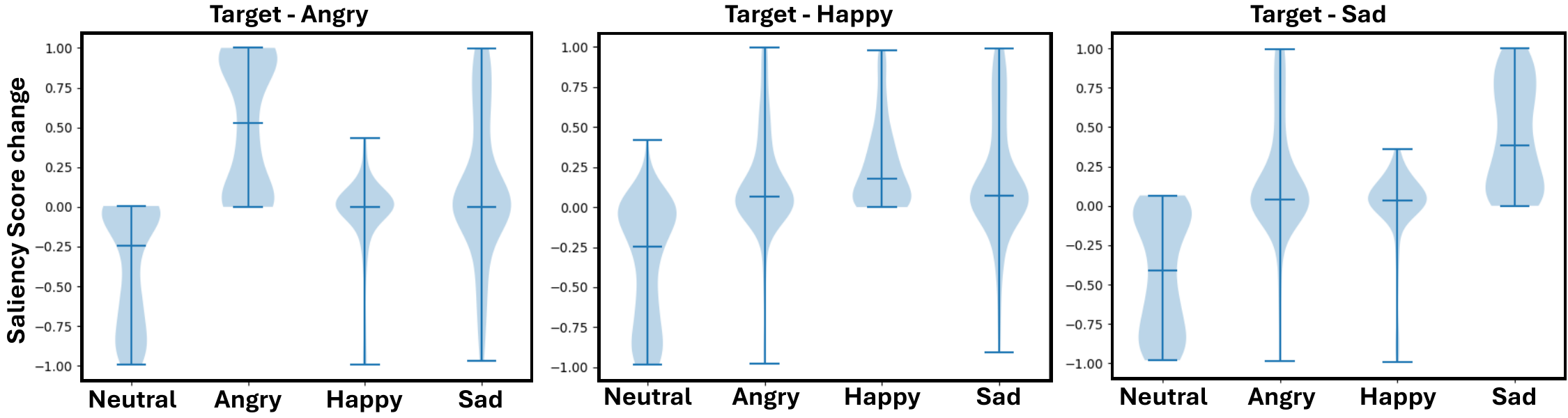}
    \caption{Absolute saliency score changes when the target emotion is: (a) angry, (b) happy, and (c) sad.}
    \label{fig:score_increase}
\end{figure*}
Additionally, we also show the performance of proposed salience predictor on another standard dataset called CREMA-D(\cite{cremad}). CREMA-D also provides a crowd-sourced emotion recognition annotation. However, the number of votes per sample are inconsistent, and it covers varying degrees of expression. 

\subsection{Emotion Recognition}
As mentioned above, we evaluate the human emotion perception prediction on the VESUS and CREMA-D corpora. The results are summarized in Table~\ref{tab:emo_recog}. We can see that the top-1 results (weighted F1 and accuracy) are above 75$\%$ for VESUS and 65$\%$ for CREMA-D. It shows that the salience prediction network (Fig.~\ref{fig:salience_network}) is reliable for predicting soft score over the emotion classes. We further evaluate top-2 accuracy of the proposed model by checking the presence of the mode of target distribution in the top-2 prediction scores. The accuracy score corresponding to this evaluation is $ > 90\%$ on VESUS and $ > 80\%$ on CREMA-D. This analysis is particularly important because, in many cases the ground-truth saliency score is a tie among two emotion categories. The top-1 prediction completely ignores this scenario whereas top-2 compensates for it. 

Fig.~\ref{fig:confusion_matrix} shows the confusion matrices on the test set obtained from proposed model. We can see that on VESUS dataset, the diagonal elements are higher for most emotions. Fear and sad categories have relatively higher confusion among themselves than any other emotion pairs. This is expected because fear and sadness are usually expressed with a shaky high-pitched articulation which can be confusing for the listeners. Further, fear has the lowest voter confidence among any category as shown in~\cite{vesus}. CREMA-D results are more muddled than VESUS due to inconsistency in number of reviewer's ratings. Here, we see that happy, sad and fear are most confused emotions with no clear pattern among them. Due to these reasons, we will focus only on VESUS corpus for the next task which is RL-based prosody manipulation using the salience predictor as feedback. 




Figure~\ref{fig:segments_demo} shows an example of the discovered segment and the corresponding prediction on a test utterances from VESUS test set. The top plot (Fig.\ref{fig:segments_demo}(a)) is the audio signal while the second plot (Fig.\ref{fig:segments_demo}(b)) is the extracted features from the convolutional encoder portion of salience predictor (Fig.~\ref{fig:salience_network}). The third plot is the variational posterior predicted (Fig.\ref{fig:segments_demo}(c)) for the masking random variable and the corresponding mask sample (in blue) obtained via sampling. The sampled mask is also overlaid on extracted features (in white). Finally, the last bar plot (Fig.\ref{fig:segments_demo}(d)) shows the ground-truth and predicted salience score over target emotion categories. Note that, the mask is one-hot in nature as it is generated via sampling from the predicted posterior. This allows the salience predictor to only look at the encodings in masked region for emotion prediction. We preserve the local property of speech by using small 1-D kernels for convolution in the feature extractor module. 

\begin{table}[t!]
\begin{center}
\begin{tabular}[t]{ |c|c|c|c|c|c| }
\hline
\textbf{Regime} & \textbf{Model} & \textbf{Angry} & \textbf{Happy} & \textbf{Sad} & \textbf{Tied}\\
\hline
\multirow{3}{3em}{Super.} & GMM-GV & 0.65 & 0.26 & 0.42 & x \\ 
& Bi-LSTM & 0.9 & 0.22 & 0.43 & x \\ 
& EDP & 0.80 & 0.63 & 0.73 & x \\ 
\hline
\multirow{3}{3em}{Unsup.} & CGAN & 0.65 & 0.20 & 0.63 & x \\ 
& VCGAN & 0.67 & 0.37 & 0.71 & x \\ 
& \textbf{Prop.} & 0.75 & 0.70 & 0.66 & \checkmark \\ 
\hline
\end{tabular}
\caption{Emotion conversion accuracy evaluated by salience predictor and compared against SOTA baselines.}
\label{tab:conv_comparison}
\end{center}
\end{table}

\subsection{Emotion Conversion}
In order to change the prosody of segments estimated by salience predictor, we train the RL agent by sampling one of the contiguous chunks during training. During inference, the individual chunks are separately processed which provides more flexibility in terms of rhythm, loudness and pitch manipulation as different segments can undergo varying degrees of modification. We evaluate the emotion conversion module on VESUS dataset for three primary categories of emotion: angry, happy and sad. Our first two evaluations use the salience prediction mode as the evaluator of emotion conversion. Specifically, we observe the positive change in score of the target emotion (Fig.~\ref{fig:score_increase}) of modified samples against state-of-the-art baselines. 
\subsubsection{Baselines}
Our baseline models are from supervised and unsupervised domain. From supervised domain, we select three models: a Gaussian mixture model (GMM) with global variance technique proposed in~\cite{gmm_emo_conv}, an Bi-LSTM model using wavelet parameterization of prosodic features proposed in~\cite{lstm_emo_conv} and a regularized convolutional neural network model (EDP) proposed in~\cite{chained_model}. The unsupervised baselines consist of cycle-GAN (CGAN) formulation using wavelet parameterization of pitch suggested in~\cite{cyclegan_emo_conv}, and the variational cycle-GAN (VCGAN) using diffeomorphic flow regularization proposed in~\cite{vcgan_journal}. 

\subsubsection{Emotion Recognition of Modified Samples}
Table~\ref{tab:conv_comparison} shows the accuracy of modified samples estimated using the salience prediction network trained via Markov masking strategy. We can observe that supervised model (EDP) has the best overall performance across all emotion categories. Our proposed technique comes very close to the supervised method showing that by carefully modeling the prosody modification, we can achieve on-par performance even in unsupervised setting. Further, our proposed model is the only technique that uses a single model for generating all target emotions unlike baselines, which train one model for each target emotion. 

Fig.~\ref{fig:score_increase} shows the change in score of the target emotions on the test set. We can see that the pattern of score difference is highest for angry followed by the sad emotion. Happy as a target emotion proves to be more challenging which can be attributed to two reasons: (a) it has the least amount of data in training set, and (b) the raters' confidence for selecting happy are lowest~\cite{vesus} among the three. Therefore, the saliency predictor can only guide the reinforcement agent in a limited capacity. It also suggests that our modeling strategy is aligned with the nuances of the VESUS dataset. Finally, our model automatically selects neutral class to increase the score for target emotion which reflects the implicit grounding of VESUS utterances in neutral emotion. Therefore, the proposed model can be used to manipulate/inject emotions into neutral utterances which is easily produced by most neural text-to-speech (TTS) synthesizers. It can also serve as a data-augmentation tool to condition TTS models on emotion for a more diverse generation. 


\subsubsection{Subjective Evaluation via A/B Testing}
\begin{figure}[!t]
    \centering
    \includegraphics[width=0.95\linewidth, height=5cm]{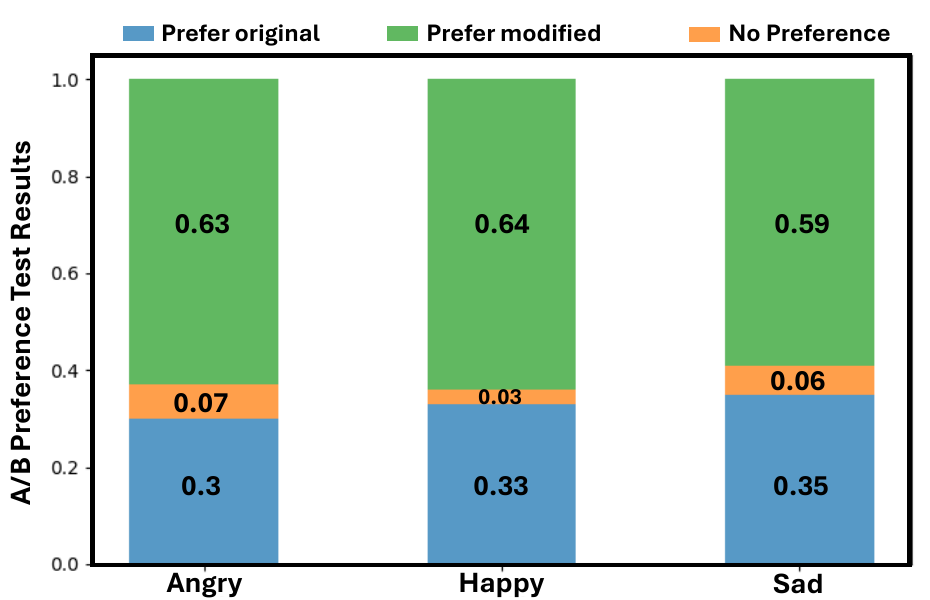}
    \caption{Preference score evaluated using crowd-sourcing random subset of test set on Amazon Mechanical Turk.}
    \label{fig:pref_score}
\end{figure}

Next, we use Amazon Mechanical Turk (AMT) platform to conduct an A/B listening test where raters are asked to pick between unmodified and modified sample for a target emotion. We also provide a no preference as third option for difficult cases. The results of this subjective listening are presented in Fig.~\ref{fig:pref_score}. We notice that in more than $50\%$ of the cases, the listeners selected the modified samples as representing target emotion. It confirms the ability of the reinforcement model to transfer or modify emotions by segment selection strategy. Additionally, we modify the emotional characteristics of the speech signal with a unified model which is advantageous in low-resource scenario. 

\subsubsection{Intelligibility Assessment}
\begin{figure}[!t]
    \centering
    \includegraphics[width=0.95\linewidth, height=5cm]{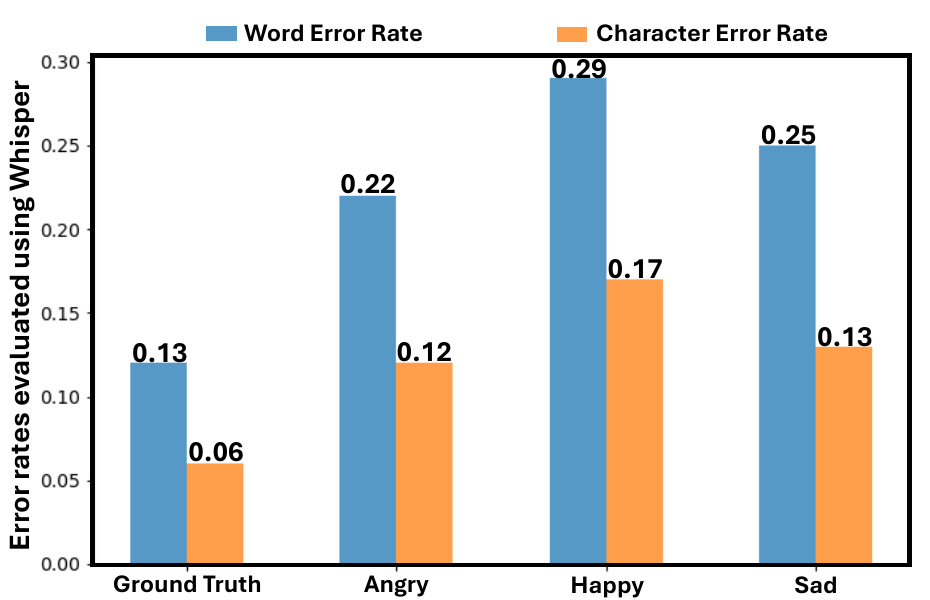}
    \caption{Intelligibility assessment of modified samples via ASR decoding using OpenAI's Whisper-medium model.}
    \label{fig:intel_score}
\end{figure}
We objectively evaluate the intelligibility of modified utterances by passing them through OpenAI's Whisper~\cite{whisper} model. We compare the word error and character error rate of modified signal using ground truth as baseline. Fig.~\ref{fig:intel_score} shows the result of this comparison. We can see that the ground-truth samples have a word error rate of $0.13$ which is comparatively lower than the modified samples in any emotion category. This suggests that prosody modification impacts the recognition performance significantly. However, this is only one part of the puzzle, the moderate increase in character error rate (compared to ground-truth) shows that certain phonemes become difficult to identify post-modification which results in poor word error. The main reason is due to extreme modifications allowed in the prosody by reinforcement agent. We can control this in a desirable manner by reducing the scales of modification. Our technique facilitates convenient tuning or control of artifacts in generated samples.

\section{Conclusion}
In conclusion, the reinforcement learning model developed for rhythm modification in speech for emotional speech synthesis represents a significant advancement in the field of emotional speech generation. By effectively identifying contiguous segments crucial for emotion perception through the innovative Markov masking strategy and implementing KL divergence-based sparsity loss, the model not only excels in emotion recognition on the VESUS corpus but also provides valuable insights into the identification of speech segments for duration modification. We show that by training the reinforcement learning agent to estimate a distribution over discrete modification factors in conjunction with WSOLA allows us to enhancement or inject emotional attributes of a speech signal. Our modeling approach is completely designed in a bottom-up generative approach with a Markov prior. We conduct both objective and subjective evaluations to further establish the effectiveness of proposed technique for emotional speech generation. We also discuss that prosody modification can lead to loss of intelligibility in the output which requires more analysis and research. Code will be made publicly available here: \url{https://github.com/ravi-0841/fac-ppg}.



\bibliographystyle{IEEEtran}
\bibliography{main}

\end{document}